\documentclass[pdflatex,sn-basic]{sn-jnl}

\usepackage{graphicx}%
\usepackage{multirow}%
\usepackage{amsmath,amssymb,amsfonts}%
\usepackage{amsthm}%
\usepackage{mathrsfs}%
\usepackage[title]{appendix}%
\usepackage{xcolor}%
\usepackage{textcomp}%
\usepackage{manyfoot}%
\usepackage{booktabs}%
\usepackage{algorithm}%
\usepackage{algorithmicx}%
\usepackage{algpseudocode}%
\usepackage{listings}%

\newcommand{\lya}{Ly$\alpha$}

\newcommand{\kms}{\rm km~s\ensuremath{^{-1}\,}}
\newcommand{\nhi}{$N$(H\,{\sc i})}

\def\ltsima{$\; \buildrel < \over \sim \;$}
\def\simlt{\lower.5ex\hbox{\ltsima}}
\def\gtsima{$\; \buildrel > \over \sim \;$}
\def\simgt{\lower.5ex\hbox{\gtsima}}

\begin{document}

\title[]{Precision Cosmology with the Lightest Elements}


\author*[1,2]{\fnm{Max} \sur{Pettini}}\email{maxpettini@icloud.com}

\author*[3]{\fnm{Ryan} \sur{Cooke}}\email{ryan.j.cooke@durham.ac.uk}
\equalcont{These authors contributed equally to this work.}

\affil*[1]{\orgdiv{Institute of Astronomy}, \orgname{University of Cambridge}, \orgaddress{\street{Madingley Road}, \city{Cambridge}, \postcode{CB3 0HA}, \country{UK}}}

\affil[2]{\orgdiv{Kavli Institute for Cosmology Cambridge}, \orgname{University of Cambridge}, \orgaddress{\street{Madingley Road}, \city{Cambridge}, \postcode{CB3 0HA}, \country{UK}}}

\affil[3]{\orgdiv{Centre for Extragalactic Astronomy, Department of Physics}, \orgname{Durham University}, \orgaddress{\street{South Road}, \city{Durham}, \postcode{DH1 3LE}, \country{UK}}}


\abstract{This is a transcript of the joint talk we gave at the Sixth Gruber Cosmology Conference at Yale University on 3 October 2025. We describe the key role played by Big Bang Nucleosynthesis (BBN) in today's `Precision Cosmology', focusing in particular on the precise determination of the primordial abundance of deuterium. We describe the development of the ideas and methods of BBN research from their beginnings more than 75 years ago to the latest developments, and conclude with  a forward look to likely advances expected towards the end of the current decade.}

\keywords{Big Bang nucleosynthesis; Cosmic microwave background radiation; Cosmological parameters; Quasar absorption line spectroscopy}



\maketitle

\section{Introduction and Historical Background}\label{Intro}

Primordial nucleosynthesis plays a unique role in cosmology: it is the only tool at our disposal to probe \textit{observationally} 
physical processes that took place very early in the history of  cosmic expansion, mere seconds and minutes after the Big Bang, and much earlier than the lifting of the curtain on the visible universe $\sim 390\,000$ years later when the cosmic microwave background (CMB) radiation was emitted.   

The early seeds of Big Bang Nucleosythesis (BBN) were sown by George Gamow, his PhD student Ralph Alpher, and Robert Herman in the years after the Second World War \citep[e.g.][]{Alpher48a}. These physicists realised that in the early stages of the universe expansion temperature and densities were sufficiently high to favour a number of nuclear reactions involving protons and neutrons leading to the synthesis of heavier elements.
Their ideas remained in embryonic form for 20 years, until the discovery of the Cosmic Background Radiation by Penzias and Wilson in 1964-65 (the CMB had incidentally been predicted in 1948 by Alpher \citep[][]{Alpher48b}, as a second natural consequence of an early hot and dense phase in the universe expansion). 
Soon after, the first quantitative calculations of the network of relevant nuclear reactions were produced by Jim Peebles
\citep[][]{Peebles66}, who had been one of the four authors of the 1965 paper interpreting the radio `noise' detected  by Penzias and Wilson as cosmic blackbody radiation, and by Robert Wagoner \citep[][]{Wagoner67}  while he was a postdoc at Caltech working with Fred Hoyle (yes, Fred Hoyle!). Figure~1 of Wagoner's paper is not very different from Figure~1 below, except that it was in black and white.

\begin{figure}[h]
\centering
\vspace*{-0.25cm}\includegraphics[width=0.75\textwidth]{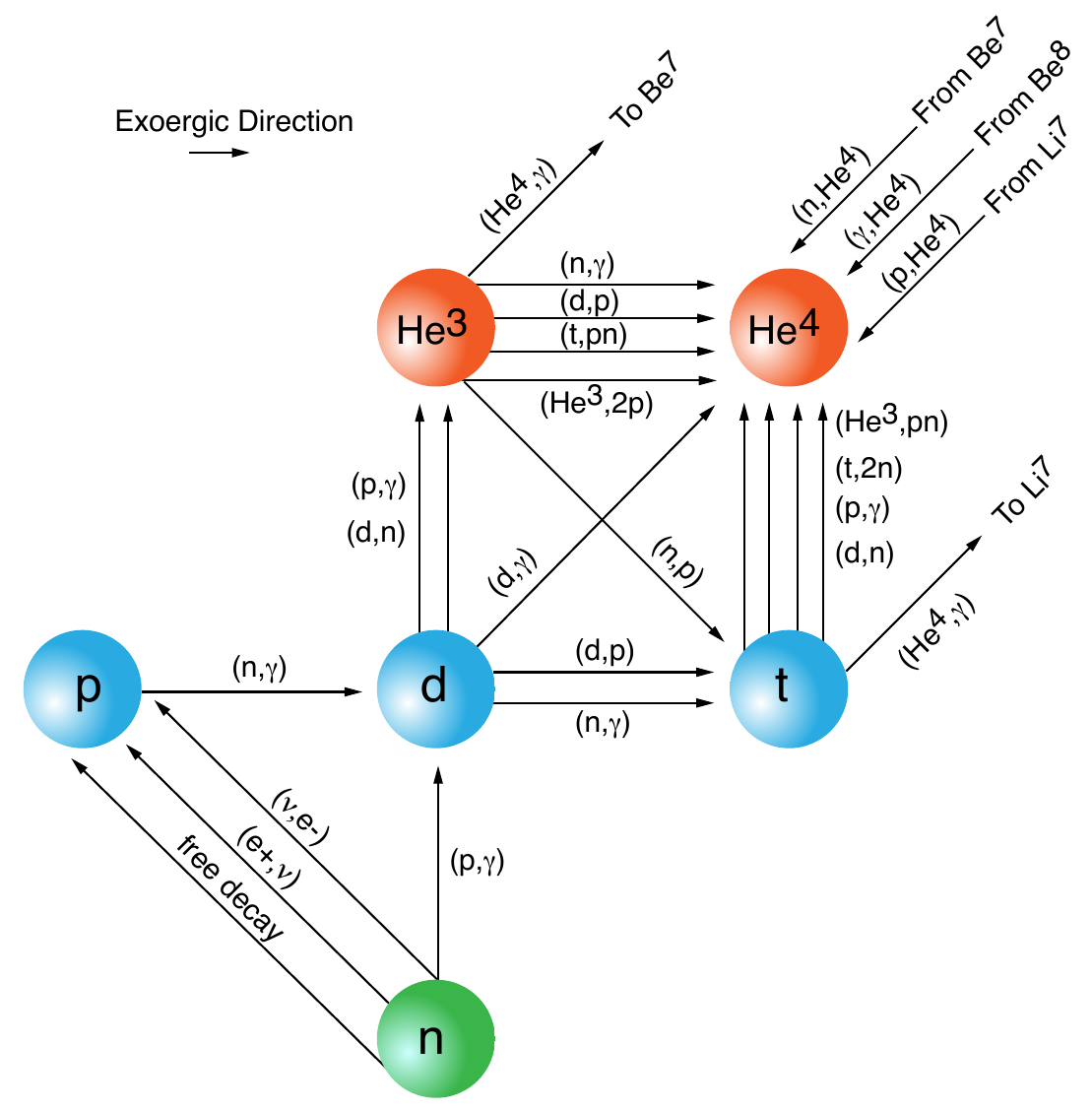}
\vspace*{0.35cm}
\caption{The nuclear reactions of Big Bang Nucleosynthesis, leading to the formation of H, He, Li and their isotopes. At the end of BBN, most neutrons are incorporated into $^4$He, with only trace amounts of deuterium, $^3$He, and $^7$Li. (Image reproduced from {\tt https://cococubed.com/} courtesy of Frank Timmes).}
\label{BBN_network}
\end{figure}

\begin{figure}[h]
\centering
\vspace*{-0.025cm}\includegraphics[width=0.85\textwidth]{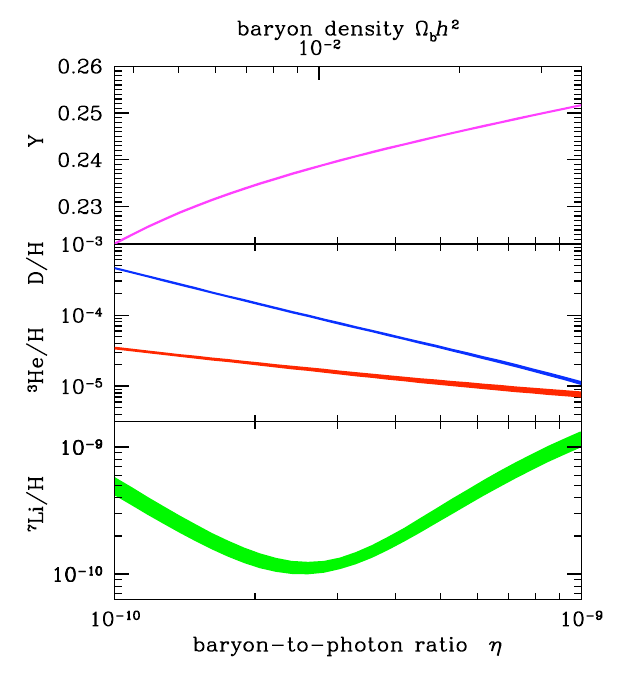}
\caption{The relative abundances of the light elements produced in BBN depend on the baryon to photon ratio, $\eta$, which in turn gives a measure of $\Omega_{\rm b} h^2$. Here $\Omega_{\rm b}$
is the fraction of the critical density contributed by ordinary matter (baryons) and $h$ is Hubble constant in units of 100\,km~s$^{-1}$~Mpc$^{-1}$. \citep[Figure reproduced from][]{Cyburt16}.}
\label{LightElem_cyburt15}
\end{figure}

Since that early work, there have been very many papers exploring BBN. Two recent reviews can be found in \cite{Fields20} and \cite{Cooke25}.
For our present purposes however, all we need to know about BBN is that in the space of a few minutes nearly all free neutrons end up in $^4$He before the expansion of the universe lowers the temperature and density to the point where the nuclear reactions involved are no longer viable. 
A much smaller proportion of the neutrons remain in isotopes of H and He, as well as Li. The exact amounts depend on the ratio of matter to light, more specifically the ratio of ordinary matter (what astronomers call baryons) to photons, or $\eta$. Since we know very accurately the number of photons from the blackbody temperature of the CMB, the primordial proportions of these light elements are a measure of the matter (baryon) content of our universe, usually expressed as a fraction of the critical density $\Omega_{\rm b}$ (see Figure~\ref{LightElem_cyburt15}).

Michael Turner and the much missed David Schramm, both from Chicago, emphasised the key role played by deuterium (D, or $^2$H) in the measurement of the baryon density of the universe. Note from Figure~\ref{LightElem_cyburt15} that the higher the density of matter, the more efficient is the conversion of D into $^4$He, and thus a lower primordial D/H ratio is left over from BBN.
The punch line here is this: measure primordial D (or any of the other light elements) with sufficient precision, and you can truly weigh the universe (in ordinary matter at least).

\section{Detecting Deuterium}\label{sec2}

So, how do we recognise and measure deuterium? The straightforward answer is in the \textit{absorption} spectra of bright background sources where each absorption line of neutral hydrogen (H\,{\sc i}) is, at least in principle, a blend of two lines, one from $^1$H and the other from $^2$H (although to be fair in most cases the $^2$H counterpart is too weak to produce detectable absorption).   
The D\,{\sc i} absorption lines are all shifted to slightly lower wavelengths, because all the electronic levels of $^2$H lie at slightly higher energies than those of $^1$H (the nucleus of $^2$H, which includes a neutron, is heavier than the nucleus of $^1$H). The isotope shift between $^1$H and $^2$H amounts to $-82$\,km~s$^{-1}$ in velocity units.

The detection of deuterium in the interstellar medium close to our location in the Galaxy was one of the first results 
to be obtained with the \textit{Copernicus} satellite, the NASA mission that literally opened the doors to UV astronomy.  \cite{RogYork73} recognised a seemingly unremarkable absorption dip in the blue wing of the interstellar Ly$\gamma$ line in the spectrum of the bright star $\beta$ Cen 
as being due to the heavy isotope of H.

The Copernicus measurement, pioneering as it was, could only give us an upper limit on the cosmic density of baryons. The reason is that all of the deuterium that exists in the universe was created in those first few minutes. But whenever stars form out of interstellar gas, some of that primordial D is burnt in the process -- this is what astronomers call the `astration' of deuterium. Consequently, only some fraction of the primordial amount remains today in the interstellar medium of the Milky Way galaxy where star formation has been progressing for more than 13 billion years. While we can in principle calculate the fraction that has been destroyed, there are several uncertainties in the calculation that preclude deducing the primordial abundance with an interesting degree of precision.

The astration of deuterium by star formation goes hand in hand with an increase in the elements of the Periodic Table that are produced by \textit{stellar} (as opposed to primordial) nucleosynthesis. To distinguish them from the light elements produced by BBN, astronomers label all elements from carbon to plutonium as `metals' (a terminology which is undoubtedly confusing to many physicists). In the interstellar medium of the Milky Way today (as in the Sun), metals account for just under 2\% of the total mass content, the remaining 98\% being H and He \citep[][]{Lodders25}. When astronomers determine the `metallicity' $Z$ of a star or a galaxy, they refer their measure to this local value.

\section{In Search of Primordial Deuterium}
We therefore have to look for astrophysical environments of much lower metallicity than $Z = 0.012$ to recover the \textit{primordial} abundance of D and other light elements produced by BBN.
In a truly prescient paper \citep[][]{Adams76}, Thomas Adams identified the numerous H absorption lines commonly seen in the spectra of distant quasars, or QSOs, as the most suitable targets in the search for primordial deuterium. Quoting from the Abstract of his paper: ``It is shown that in suitable QSO absorbing clouds the deuterium Lyman alpha line should be detectable... Observers should be alert for the deuterium Lyman alpha line since its detection would have important cosmological implications''.

\begin{figure}[h]
\centering
\vspace*{-0.025cm}\includegraphics[width=0.95\textwidth]{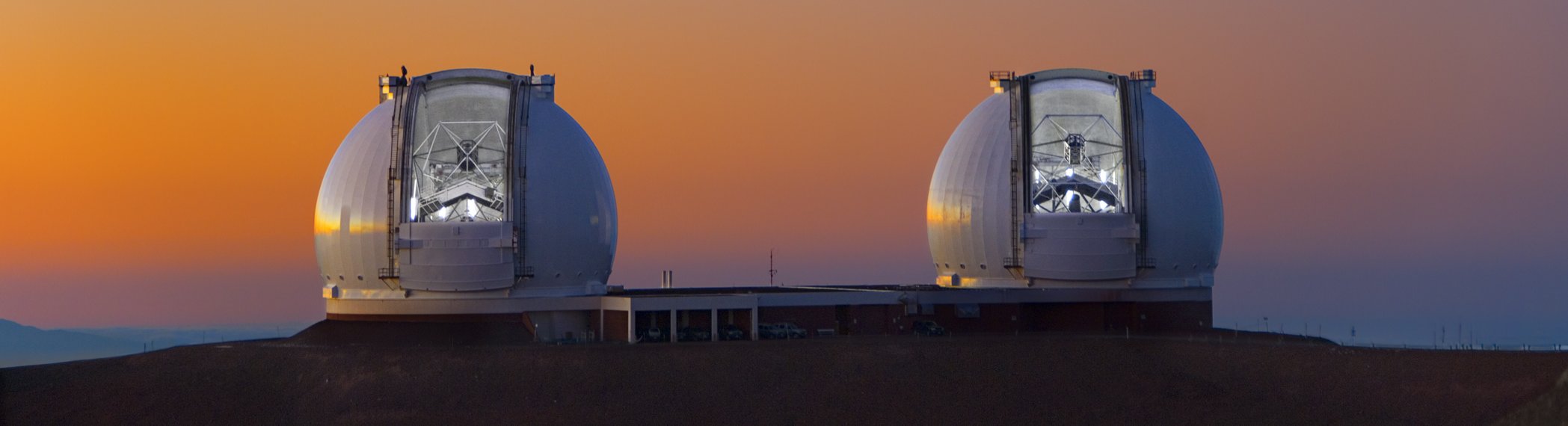}
\caption{The twin 10-m Keck telescopes on the summit of Manua Kea, on the Big Island of Hawai'i.}
\label{KeckTels}
\end{figure}

We had to wait twenty years before the advent of astronomical instrumentation capable of turning Adams' vision into reality. David Tytler \citep[][]{Tytler96} led the way with the first successful detection of D absorption in quasar spectra using the newly commissioned HIRES spectrograph on the Keck I telescope (Figure~\ref{KeckTels}). Tytler and collaborators targeted a low metallicity ($Z \simeq 1/100 Z_\odot$) gas cloud at redshift $z_{\rm abs} = 3.572$ with a neutral hydrogen column density $\log_{10} (N$(H\,{\sc i})/cm$^{-2}) = 17.94$, seen against the background light of the bright quasar 1937$-$1009 at $z_{\rm em} = 3.78$. Their HIRES spectrum clearly showed D\,{\sc i} absorption in the blue wings of the two strongest lines in the Lyman series, Ly$\alpha$ and Ly$\beta$, from which the authors deduced the deuterium abundance: ${\rm (D/H)} = (2.3 \pm 0.6) \times 10^{-5}$.
At $Z = 1/100 Z_\odot$ the astration of D would have been minimal, so this value is taken to be the primordial abundance of deuterium.

\section{Precision Measures of (D/H)$_{\rm p}$}

With $N$(H\,{\sc i})~$\simeq 10^{18}$\,cm$^{-2}$, the absorbing cloud selected by Tytler and colleagues belongs to a class of QSO absorbers termed `Lyman Lymit Systems' (LLS) because they produce a discontinuity in the quasar spectrum at the (redshifted) wavelength of the limit of the Lyman series of H\,{\sc i}, corresponding to a rest frame wavelength of 912\,\AA---photons with this energy (13.6\,eV) readily ionise H\,{\sc i} and are easily absorbed by gas with $N$(H\,{\sc i})$ \simgt 3 \times 10^{17}$\,cm$^{-2}$.

\begin{figure}[h]
\centering
\hspace*{-0.25cm}\includegraphics[width=0.8\textwidth]{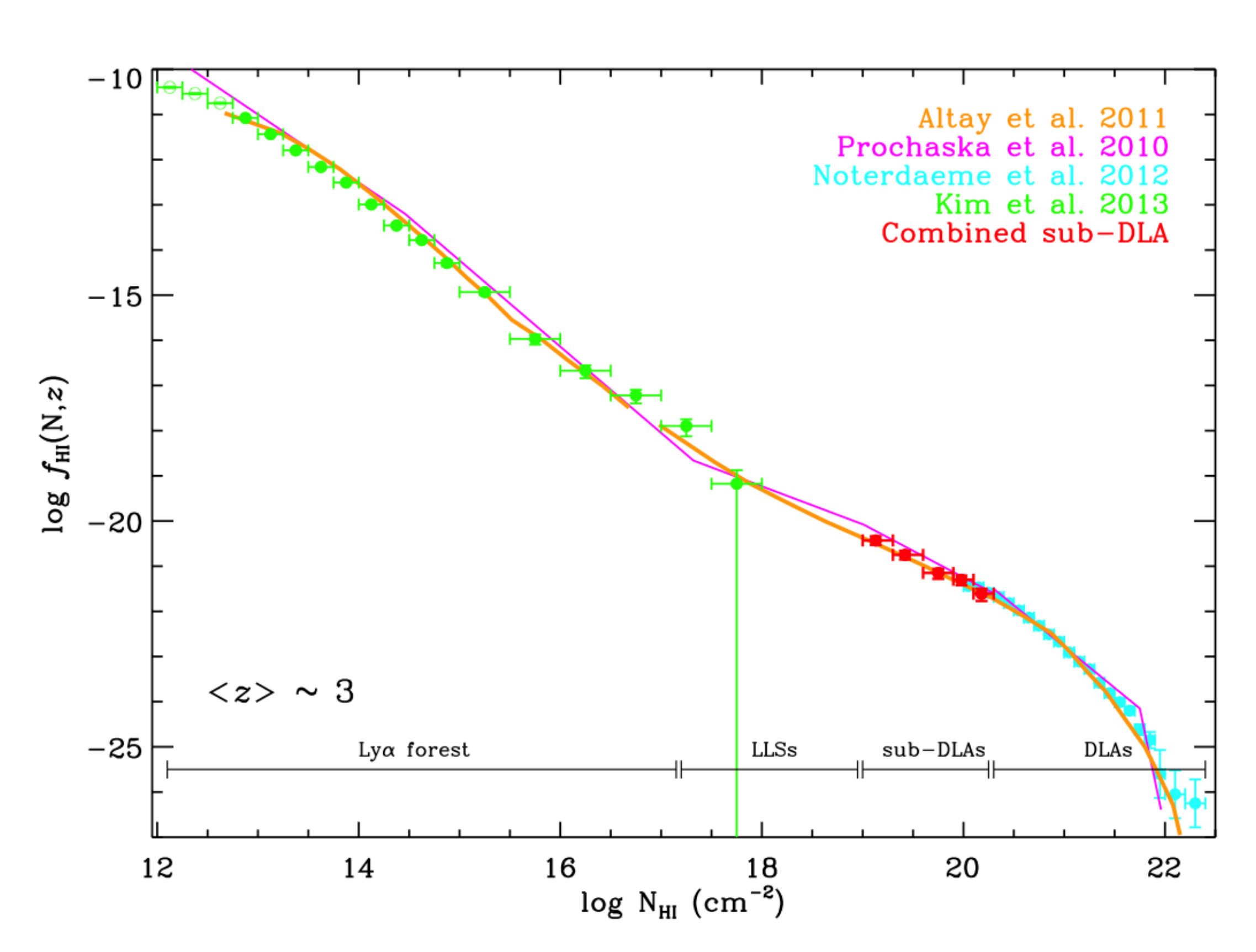}
\caption{Column density distribution of H\,{\sc i} absorbers \citep[reproduced from][]{Zafar13}.}
\label{fig:N(HI)}
\end{figure}

LLSs are but one subset of a distribution of H\,{\sc i} absorbing clouds that spans ten orders of magnitude (see Figure~\ref{fig:N(HI)}),
from $N$(H\,{\sc i})\,$\simeq 10^{12}$\,cm$^{-2}$ --- wisps of neutral gas in an otherwise highly ionised intergalactic medium --- to the `heavy weights' at $N$(H\,{\sc i})\,$\simeq 10^{22}$\,cm$^{-2}$.
The latter were identified by Art Wolfe some forty years ago as a particularly interesting class of QSO absorbers, likely associated with the inner, neutral interstellar medium of galaxies
\citep[see][for a review]{Wolfe05}. Wolfe coined the term `damped Lyman alpha systems' (DLAs), to indicate systems with column densities $N$(H\,{\sc i})\,$ > 2 \times 10^{20}$\,cm$^{-2}$.

It can be readily appreciated from Figure~\ref{fig:N(HI)} that DLAs
are the rarest of QSO absorbers. Despite their rarity, several tens of thousand of DLAs (some candidates and some confirmed) are now known, thanks to large scale surveys of the sky, such as the \textit{Sloan Digital Sky Survey, SDSS} --- the brainchild of Jim Gunn, the 2005 Gruber Cosmology Laureate. With such high values of $N$(H\,{\sc i}), essentially all known DLAs also exhibit many absorption lines from metals, allowing accurate assessment of their chemical composition (i.e. the relative proportions of different elements) and, crucially for our present purposes, the overall metallicity (the ratio of metals to hydrogen). 

DLAs are a mixed bunch when it comes to their metallicity. The majority are metal-poor, with only $\sim 1/20$--$1/30$ of the metals found in the Sun and in the ISM of the Milky Way today. However, the distribution of $Z_{\rm DLA}$ values spans (at least) three orders of magnitudes, possibly reflecting the wide range of galaxy types (and masses) capable of producing a damped Ly$\alpha$ line when viewed in absorption \citep[][]{Pontzen08}. One tail of the distribution stretches to near-solar metallicities, while at the other end several DLAs are known with metallicities $Z_{\rm DLA} \leq 1/1000 \, Z_\odot$ \citep[e.g.][]{Welsh24}.

\subsection{Primordial Deuterium in Metal-poor Damped Lyman $\alpha$ Systems}

\begin{figure}[b!]
\centering
\hspace*{-0.025cm}\includegraphics[width=0.8\textwidth]{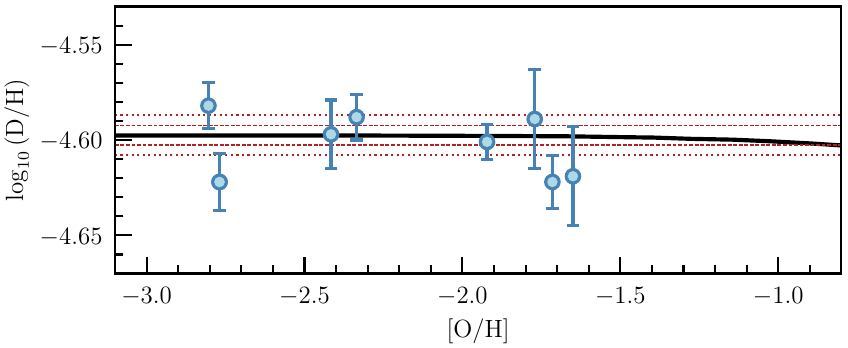}
\caption{The `precision sample' of eight measures of the D/H in metal-poor DLAs, plotted against the oxygen abundance. $1\,\sigma$ errors are shown. Seven of the values  are from \cite{Cooke18}, while the eighth is from the recent work by \cite{Kislitsyn24}.  The dotted lines indicate the 1 and 2\,$\sigma$ bounds on the weighted mean: $\langle {\rm D/H} \rangle = (2.510 \pm 0.028) \times 10^{-5}$. The black line 
shows the astration of deuterium estimated from the chemical evolution model of \cite{Voort18}: for DLAs with [O/H]\,$\simlt -1.5$ the correction for astration is much smaller than the errors on the D/H measurements.
}
\label{fig:PrecSample}
\end{figure}

\begin{figure}[h]
\centering
\vspace*{0.25cm}\includegraphics[width=0.75\textwidth]{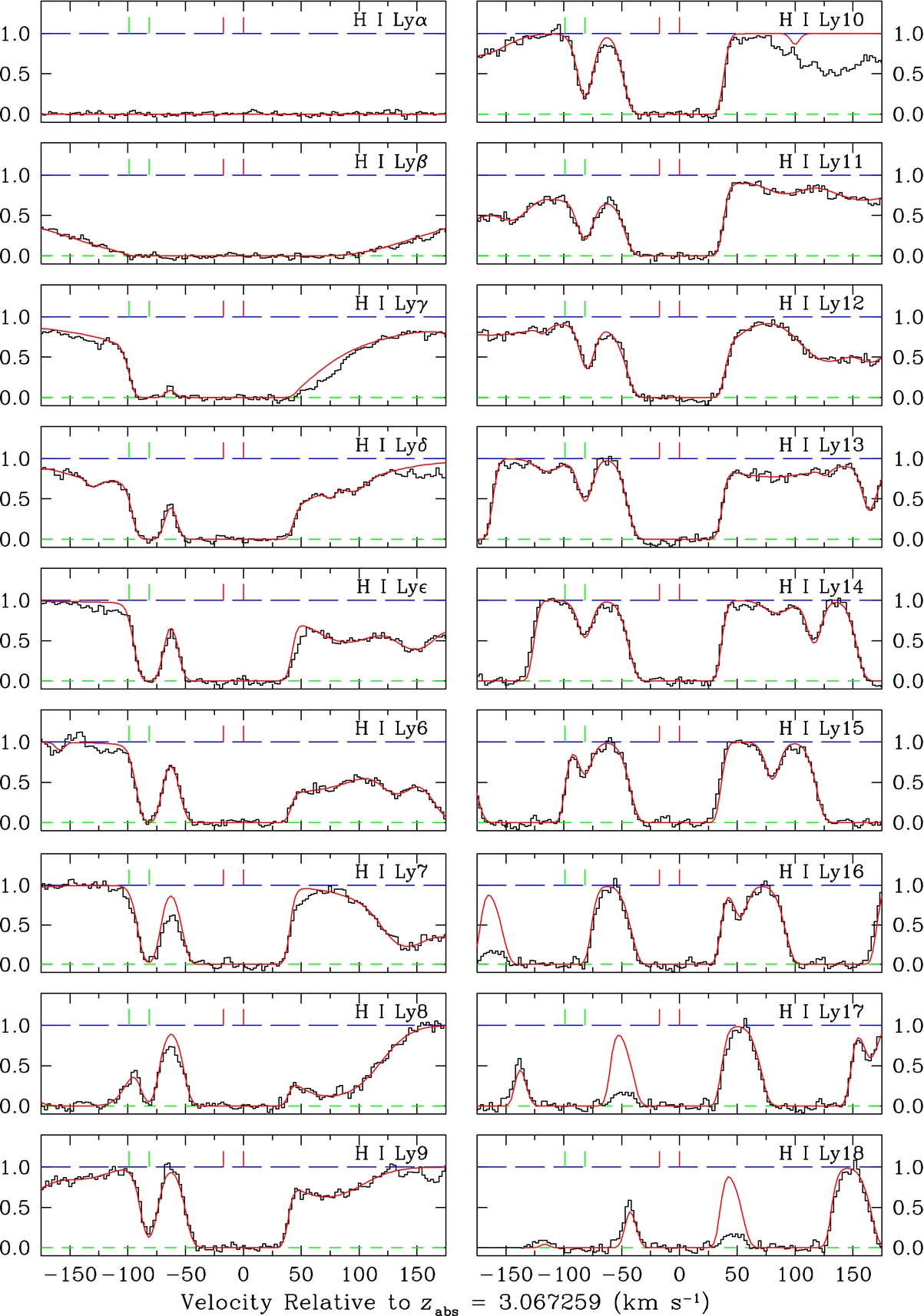}
\medskip
\caption{Absorption by D\,{\sc i} and H\,{\sc i} absorption lines (indicated by the green and red ticks above each spectrum, respectively) in the Lyman series of the $z_{\rm abs} = 3.0673$ DLA seen in the spectrum of the QSO J1358+6522.
D\,{\sc i} absorption is resolved from Ly$\gamma$ to Ly15 \citep[reproduced from][]{Cooke14}.}
\label{fig:J1358}
\end{figure}

It is these most metal-poor DLAs that we identified as the most suitable astrophysical environments for measuring the primordial abundance of deuterium for the following reasons:

\begin{itemize}
\item The low metallicities ensure minimal astration of deuterium: the value of D/H we measure is, within the error of measurement, the primordial value (D/H)$_{\rm p}$.

\item The most metal-poor DLAs are also the ones with the lowest velocity dispersion of the absorbing gas, as expected if a mass-metallicity relation applies to their host galaxies. Indeed it has been suggested that these DLAs may be the high-redshift analogues of Local Group dwarf galaxies \citep[][]{Cooke15}. Typically, their absorption lines have widths of only a few \kms\ --- see for example Figure~8 of \cite{Cooke15}. This is important: with such low internal velocity dispersions, the 82\,\kms\  isotope shift of deuterium is usually well resolved and free from contaminating features.

\item The high \nhi\ of DLAs brings within reach \textit{many} transitions of the Lyman series, rather than just the first few as is the case for LLSs; in the best examples (e.g. Figure~\ref{fig:J1358}) we see D\,{\sc i} absorption up to Ly15 (!), providing us with several independent measures, within the same spectrum, of the D\,{\sc i}/H\,{\sc i} ratio. 

\end{itemize}

The combination of parameters necessary for the precise measurement of (D/H)$_{\rm p}$ is rare. This is a delicate experiment that requires everything to be just right:
the redshift and brightness of the quasar as well as the redshift, column density, metallicity and kinematics of the damped \lya\ system --- these and other factors all conspire to narrow significantly the available parameter space suitable for a precise determination of  (D/H)$_{\rm p}$.
This goes some way to explain why, after a decade of dedicated and time-consuming observations, the sample of high-precision D/H measurements amounts to a grand total of eight sight-lines, collected in Figure~\ref{fig:PrecSample}.

There are several things to note in Figure~\ref{fig:PrecSample}. First, all eight measures of D/H are consistent with one another within the errors --- this was not always the case in earlier work \citep[see for example][]{Steigman01}.
This internal consistency was achieved in part with the high S/N of the high resolution (echelle) spectra obtained, but also by developing dedicated spectral analysis tools that take into account not only the random errors but also various sources of systematic uncertainty such as the exact placement of the continuum and background levels and the presence of unrelated absorption lines in the spectral regions of interest. In our own analyses, we chose to blind ourselves from the fitted D/H value until the data reduction and spectral fitting had reached what we considered to be an optimal fit to the data. Only then did we reveal the best fit D/H value, and subsequently refrained from any further tweaks. This procedure was adopted to reduce as much as possible the impact of human bias.

The black line in Figure~\ref{fig:PrecSample} shows the output of the galactic chemical evolution calculations by \cite{Voort18} which include a measure of the degree of astration of deuterium as a function of oxygen abundance. Evidently, the correction for the consumption of D through star formation is minimal at metallicities less than $\sim 1/30$ of solar (equivalent to $[{\rm O/H}] \lesssim -1.5$). Thus, it is reasonable to adopt the weighted mean of all eight measurements,  $\langle {\rm D/H} \rangle = (2.510 \pm 0.028) \times 10^{-5}$, as the primordial abundance of D, (D/H)$_{\rm p}$.

\section{From (D/H)$_{\rm p}$ to $\Omega_{\rm b}$}

\begin{figure}[h!]
\centering
\hspace*{-0.75cm}\includegraphics[width=0.85\textwidth]{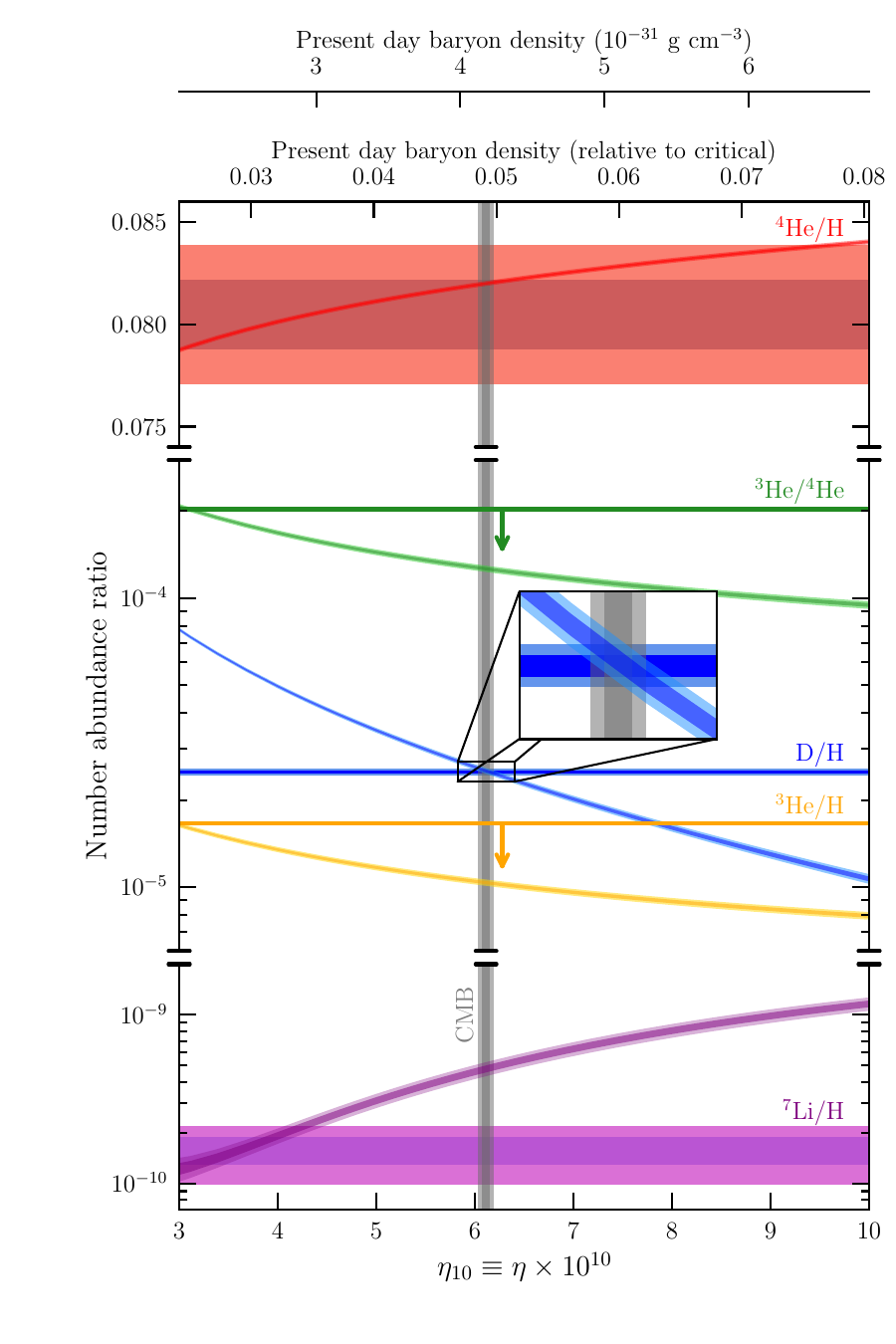}
\caption{Confronting measurements of light element abundances (coloured horizontal bands with shading to indicate 1 and 2\,$\sigma$ ranges) with expectations from the CMB value of $\Omega_{\rm b}$ (vertical grey band). The curves show the dependence of the five primordial number abundance ratios on $\eta_{10}$. There is spectacular agreement, within the small uncertainties, between $\Omega_{\rm b}$ deduced from the primordial abundance of deuterium (the `baryometer' of choice, quoting David Schramm), and from the temperature fluctuations of the CMB \citep[Figure reproduced from][]{Cooke25}.}
\label{fig:Omega_b}
\end{figure}

A $\sim 1\% $ measure of (D/H)$_{\rm p}$ can only lead to a similarly precise determination of 
$\Omega_{\rm b}$ if the cross-sections of the nuclear reaction network shown in Figure~\ref{BBN_network} are also well known. In the past, not all the channels shown in the Figure have been measured, or calculated, with the same degree of precision, but it is encouraging now to see that the recent improvements in the astrophysics have stimulated renewed interest in the relevant nuclear physics. Thus it was that a chance conversation at the 12th International Symposium on Nuclei in the Cosmos (NIC 2012), held in August 2012 in Cairns Australia, between one of us (MP) and Carlo Gustavino of the Italian Istituto Nazionale di Fisica Nucleare, motivated the 
Laboratory for Underground Nuclear Astrophysics (LUNA) located deep underground in the Apennines mountains of central Italy, to conduct a modern measurement of the $S$ factor of one of the key reactions: the fusion of D and a proton to form $^3$He with the emission of a photon---D(p,$\gamma$)$^3$He (see Figure~\ref{BBN_network}).

There remain minor differences in the way different groups interpolate and extrapolate the $S$ factors of BBN nuclear reactions in the relevant energy range, but it is hoped that the situation will improve further in future, as more channels in Figure~\ref{BBN_network} are measured with modern equipment \cite[e.g.][]{Shen24}.
Using Cara Giovanetti's new public BBN code LINX \citep[][]{Gio25},  
(D/H)$_{\rm p} = (2.510 \pm 0.028) \times 10^{-5}$ implies $\Omega_{\rm b} h^2 = 0.02241 \pm 0.00031$, adopting the PArthENoPE rates \citep{Gariazzo22}.

It is now well known that encoded in the angular power spectrum of the temperature (and polarisation) fluctuations of the Cosmic Microwave Background are virtually all cosmological parameters of interest, including $\Omega_{\rm b}$. Successive space and ground-based experiments have led to today's `Precision Cosmology' with the parameters describing our universe now known with exquisite precision (to better, and in some cases much better, than a few percent, which would have been undreamed of just a few decades ago). Turning specifically to $\Omega_{\rm b}$, by considering together data from the 
\textit{Planck} satellite and from Data Release 6 maps made with the Atacama Cosmology Telescope (ACT), \cite{ATC25} find $\Omega_{\rm b} h^2 = 0.02250 \pm 0.00011$, in excellent agreement with the value deduced from the primordial abundance of deuterium, as shown in Figure~\ref{fig:Omega_b}.

Let us reflect on this for a minute. We have pursued two entirely different avenues to determining the cosmic density of baryons. One uses a network of nuclear reactions that synthesised the lighter elements of the Periodic Table in the first few minutes of the universal expansion. The other relies on modelling acoustic oscillations in the photon-baryon fluid some 390\,000 years later. Very different physics and widely separated cosmic epochs give us  \textit{the same answer within 1\%}. This is surely a spectacular success of the Standard Model of Big Bang cosmology, a triumph worth shouting about from the roof tops.

\section{Beyond the Standard Model?}

So, why is it that astronomers are not yet done with cosmology? The truth is that this accomplishment has a bitter-sweet taste: yes, the cosmic density of ordinary matter has been measured very precisely, but the answer we have arrived at is that the matter that makes up our world accounts for just less than 5\% \footnote{Adopting $h = 0.676$ from \cite{ATC25}}
of the total energy budget of the universe. The remaining 95\% remains mysterious: there is no room in the Standard Model for either Dark Matter nor Dark Energy.

And so it is that, despite the successes of the $\Lambda$CDM model, astronomers have turned their attention to looking for `chinks in its armour' --- discrepancies in the measurements of cosmological parameters at different epochs and in different physical environments. The most visible of these is the current debate on whether the value of the Hubble constant deduced from the CMB and from Baryon Acoustic Oscillations (BAO) is significantly different from that derived from the Hubble diagrams of supernovae and other standard candles \cite[see][for an extensive review]{DiValentino25}.

Turning to the abundances of the light elements, the focus has been on whether the current `best' determinations of the primordial abundances of $^4$He, $^3$He and $^2$H
are consistent among themselves and with the CMB value of $\Omega_{\rm b}$. One form that such tests have taken is to assess whether there are significant departures from the 
`effective number of neutrino species' $N_{\rm eff} = 3.044$, appropriate to the  Standard Model of Particle Physics. The label is somewhat misleading in that $N_{\rm eff}$ is in reality just a measure of the energy content of the early universe that determines its expansion rate at the time of BBN. Nevertheless, much attention has been focused on hypothetical `sterile neutrinos', or other `dark' relativistic components, motivated by the discovery of oscillations between `active' neutrinos. Such oscillations  require both neutrinos non-zero masses and flavour mixing, whereas neither are entertained within the Standard Model.

\begin{figure}[h!]
\centering
\hspace*{-0.25cm}\includegraphics[width=0.6\textwidth]{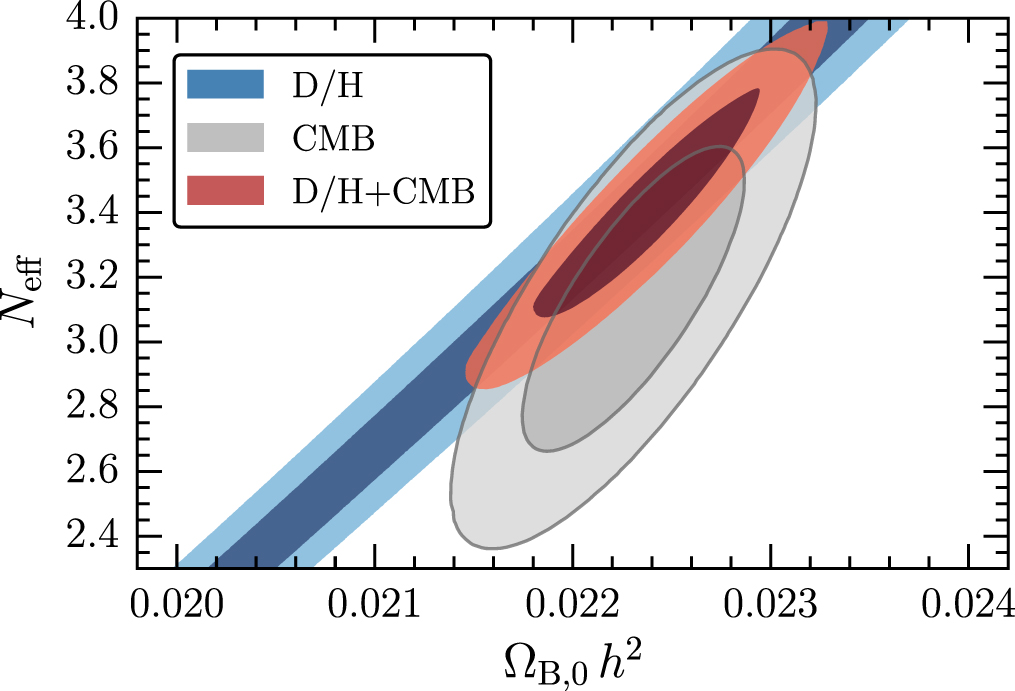}
\caption{Comparing the expansion rate (parameterized by $N_{\rm eff}$) and the cosmic density of baryons ($\Omega_{\rm b}h^2$) from BBN (blue) and CMB (grey). The red contours indicate the combined confidence bounds; dark and light shades are the 68\% and 95\% contours respectively \citep[reproduced from][]{Cooke18}.}
\label{fig:N_eff}
\end{figure}

There have been many checks on $N_{\rm eff}$ in the last decade or so, with increasing sensitivity. The bottom line is that all estimates to date are consistent with 
$N_{\rm eff} = 3.044$. One example (by no means the most discriminating) is shown in Figure~\ref{fig:N_eff} which matches (D/H)$_{\rm p}$ to the CMB; together they yield 
$N_{\rm eff} = 3.41 \pm 0.45$ \citep[][]{Cooke18}.

Despite the lack of evidence so far, there are strong incentives to push the precision 
of light element abundances beyond the current 1\% limit. As emphasised by Cara Giovanetti at this Gruber Cosmology Conference, much lies in wait below the 1\% precision in terms of differentiating between, for example, different dark matter candidates proposed. Fortunately, it is easy to identify several developments in the next decade that would allow us to push the precision of measures of (D/H)$_{\rm p}$, and other light elements, to new heights.

Observationally, the two main game changers are the advent of 30+m optical/infrared telescopes and of UV sensitive instrumentation. The European Southern Observatory have plans well under way for both: the Extremely Large Telescope (ELT) with an aperture of 39\,m, and the Cassegrain U-Band Efficient Spectrograph (CUBES) to be mounted on Unit Telescope 1 of the Very Large Telescope facility. Both are expected to come on line towards the end of the current decade or early in the 2030s. Together, these innovations will increase by at least an order of magnitude the number of QSO sight-lines accessible for sensitive measurements of (D/H)$_{\rm p}$. When coupled with projected improvements in the determinations of the cross-sections of BBN nuclear reactions, the outlook for sub-percent precision in primordial light element abundances is very positive. Nearly a century after the pioneering ideas of George Gamow and Ralph Alpher, BBN continues to be at the forefront of modern cosmology research.

\section{Conclusions}\label{sec13}

We hope that at the end of this talk you will take away the following key points:
\begin{itemize}
    \item BBN is a unique tool allowing us to probe physics 
in the first few minutes of the universe expansion.

    \item Metal-poor DLAs are the most suitable environments for 
measuring the primordial abundance of deuterium.

    \item The best sample of measures of the primordial abundance of deuterium, (D/H)$_{\rm p} = (2.510 \pm 0.028) \times 10^{-5}$,  has 
allowed us to infer the baryon density with 1\% precision:
$\Omega_{\rm b}h^2 = 0.02241 \pm 0.00031$.

    \item This value is in spectacular agreement with that deduced 
from the analysis of the temperature fluctuations of the CMB. The fact that the two methods, which probe entirely different physics and at two widely separated cosmic epochs, give such concordant answers is an emphatic validation of the Big Bang model. 

\item Up to now, convincing evidence for departures from the Standard Model as parameterised, for example, by the effective number of neutrino species, $N_{\rm eff}$, still eludes us.

\item However, looking ahead, we foresee several opportunities from improvements in: (i) the precision of (D/H)$_{\rm p}$ measures through significantly larger samples of metal-poor DLAs
made accessible to observations by new instruments such as ESO's Extremely Large Telescope and the Cassegrain U-Band Efficient Spectrograph on the Very Large Telescope; and (ii) the determination of BBN nuclear reaction rates motivated by the increasing precision in the measurements of the primordial abundances of the light elements. 

\end{itemize}

\backmatter

\bmhead{Acknowledgements}

We thank the Peter and Patricia Gruber Foundation for their generous continuing support of science through their International Prize Programs and Young Scientist Awards, and  the members of the Cosmology Selection Advisory Board whose job is made difficult, as well as pleasant, by the fast pace of advance of discoveries about our universe. 
Special thanks are due to Viatcheslav (Slava) Mukhanov. We are grateful to the Astronomy Department of Yale University for organising and hosting the Sixth Gruber Cosmology Conference at Yale University, and to the Institute of Astronomy of the University of Cambridge for providing the stimulating intellectual environment from which both of us have benefited during our careers. The work described here was carried out over many years with the help of several colleagues: David Bowen, Regina Jorgenson, Antony Lewis, Michael Murphy, Ken Nollett and Chuck Steidel: they all made vital contributions towards achieving high precision measures of the deuterium abundance.

\bibliography{BBN-bibliography}

\section*{Statements \& Declarations}

\begin{itemize}
\item The authors declare that no funds, grants, or other support were received during the preparation of this manuscript.

\item The authors have no relevant financial or non-financial interests to disclose.

\end{itemize}

\end{document}